\documentclass[
    ,final            
  ]
  {aipproc}

\layoutstyle{6x9}

\begin{document}

\title{Coherent Power Corrections to \\ Structure Functions}


\author{Ivan Vitev}{
  address={Los Alamos National Laboratory, Theory Division and 
Physics Division, Mail Stop H846, \\ Los Alamos, NM 87544}
}

\begin{abstract}

We calculate and resum a perturbative expansion of nuclear enhanced
power corrections to the structure functions measured in deeply 
inelastic scattering of leptons on a nuclear target. Our results 
for the Bjorken $x$-, $Q^2$- and $A$-dependence of nuclear shadowing 
in $F_2^A(x,Q^2)$ and the nuclear modifications to $F_L^A(x,Q^2)$, 
obtained  in terms of the QCD factorization approach, are consistent 
with the existing data.  We predict the dynamical 
shadowing from final state interactions in $\nu + A$ reactions 
for sea and valence quarks in the  structure functions $F_2^A(x,Q^2)$ 
and $x F_3^A(x,Q^2)$, respectively.
In $p+A$ collisions we calculate the centrality and rapidity 
dependent nuclear suppression of single and double inclusive 
hadron production at moderate transverse momenta.  

\end{abstract}

\maketitle

\subsection{Dynamical high twist shadowing} 

Under the approximation 
of one-photon exchange, the lepton-hadron
DIS cross section $d\sigma_{\ell h}/dx\, dQ^2 \propto L_{\mu\nu}\, 
W^{\mu\nu}(x,Q^2)$, with Bjorken variable $x=Q^2/(2p\cdot q)$ and
virtual photon's invariant mass $q^2=-Q^2$. 
The hadronic tensor can be expressed in
terms of  structure functions based on the polarization states of
the exchange virtual photon: 
$W^{\mu \nu}(x,Q^2) = \epsilon_T^{\mu \nu}\, F_T(x,Q^2) +  
\epsilon_L^{\mu \nu}\, F_L(x,Q^2)$.  In DIS the exchange 
photon  $\gamma^*$ 
of virtuality $Q^2$ and energy $\nu = Q^2/(2 x m_N)$  probes an
effective volume of transverse area $1/Q^2$ and longitudinal extent 
$\Delta z_N \times x_N/x$, where $\Delta z_N$ is the nucleon size, 
$x_N=1/(2 r_0 m_N) \sim 0.1$ 
and $r_0\sim 1.2$~fm. When Bjorken $x \ll x_N $ 
the lepton-nucleus DIS covers  several nucleons in longitudinal 
direction while it is localized in the transverse plane.

\begin{center}

\begin{figure}[!t]
  \includegraphics[height=4.in]{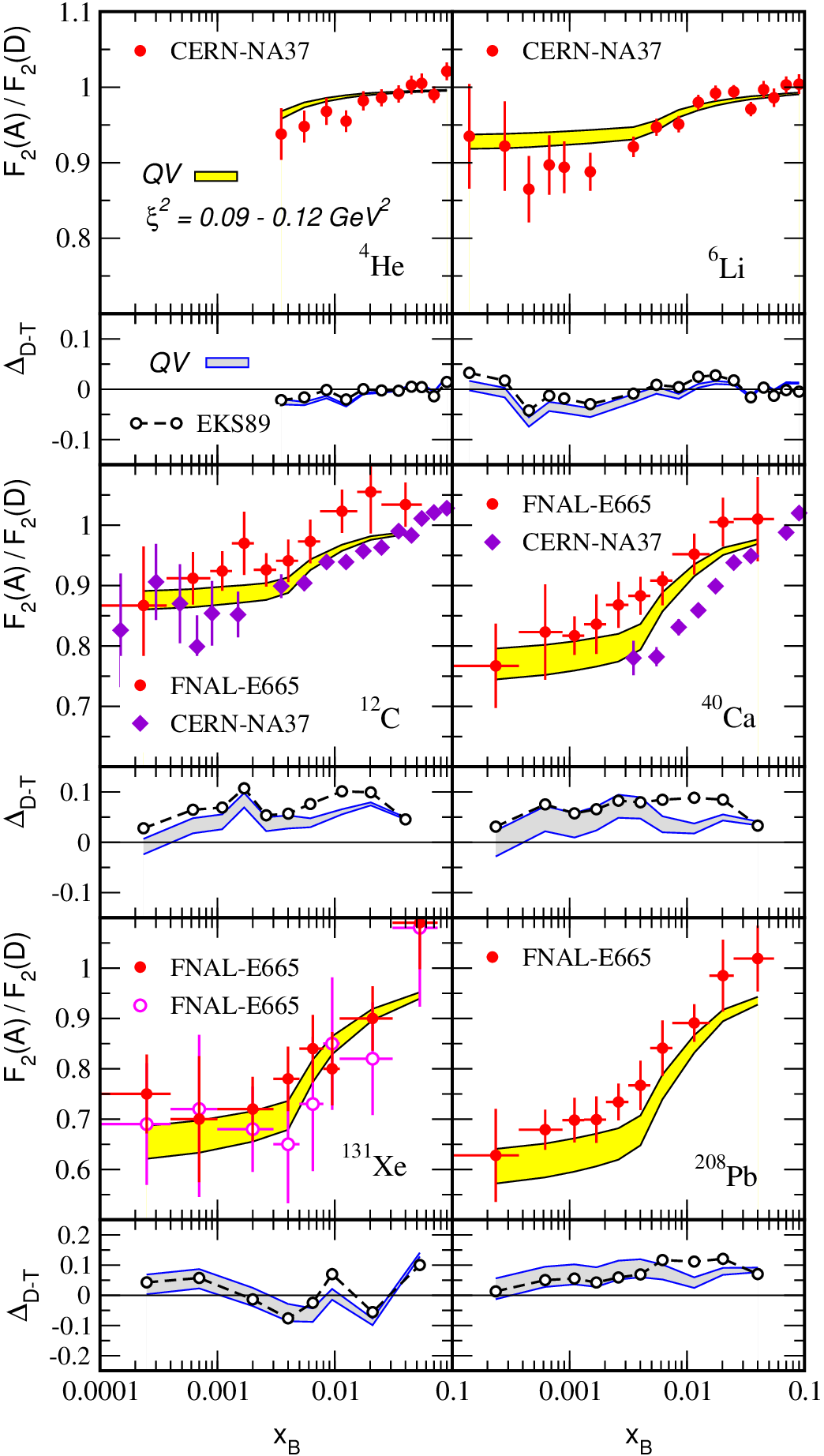} \hspace*{1cm}
  \includegraphics[height=3.in]{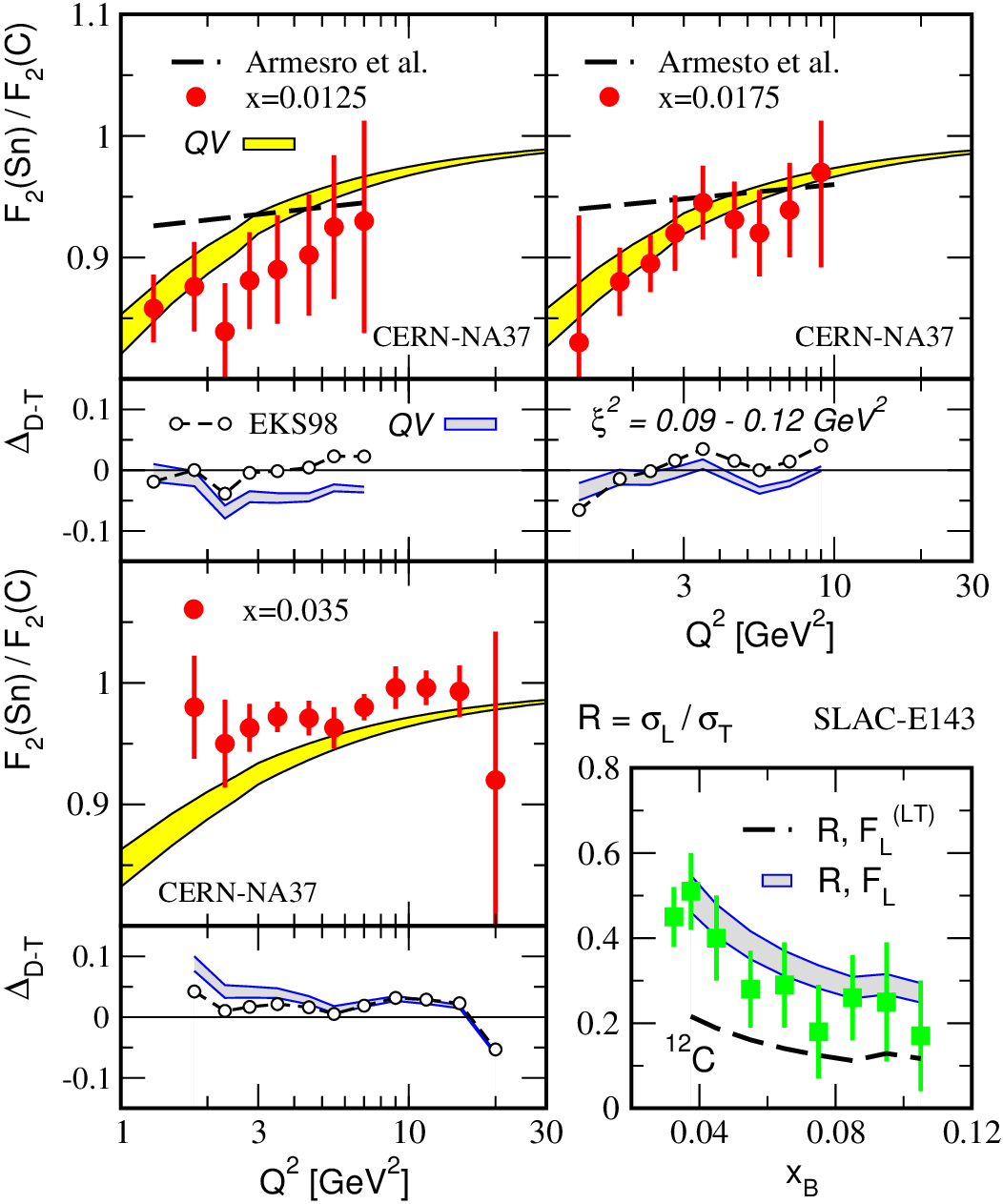}
  \caption{Left panel:  $F_2(A)/F_2(D)$ calculation 
of resummed power corrections 
versus nuclear $A$ and Bjorken-$x$~\cite{Qiu:2003vd}. 
Right panel: $F_2(Sn)/F_2(C)$ show evidence 
for a power-law  in $1/Q^2$ behavior consistent with 
the all-twist resummed calculation~\cite{Qiu:2003vd}. 
The bottom right panel illustrates the role of higher 
twist contribution to $F_L$  on the example of 
$R=\sigma_L/\sigma_T$.   } 
\end{figure}

\end{center}

In the lightcone $A^+ = 0$ gauge and the Breit frame we 
identify the natural short and long distance separation of the 
multiple final state interactions from the propagator structure of the 
struck quark,  $i(\gamma^+/2p^+)/(x_i-x\pm i\epsilon)$ (pole term) 
and $i(xp^+/Q^2)\gamma^-$ (contact term)~\cite{Qiu:2003vd}.
The two gluon  contact exchange is therefore evaluated in a single 
nucleon state.  Resumming the $A^{1/3}$-enhanced power corrections
we find~\cite{Qiu:2003vd}: 
\begin{eqnarray} 
F_T^A(x,Q^2) \! & \approx  &
A \, F_T^{\rm (LT)}\left( x + \frac{x \xi^2 ( A^{1/3}-1) }{Q^2}, 
                      Q^2 \right) \, ,
\label{FTres}  \\ 
F_L^A(x,Q^2)  \! & \approx & 
A\,  F_L^{\rm (LT)}(x,Q^2) + 
\frac{ 4\, \xi^2 }{Q^2} \, F_T^A(x,Q^2)  \; .
\label{FLres}  
\end{eqnarray}
Here, $\xi^2$ represents the characteristic scale of higer twist 
per nucleon to ${\cal O}(\alpha_s)$: 
\begin{eqnarray}
\xi^2  & = &  \frac{3 \pi \alpha_s(Q^2)}{8\, r_0^2} 
\langle p| \, \hat{F}^2 (\lambda_i) \,| p \rangle  \; , 
\qquad  
\langle p| \, \hat{F}^2 (\lambda_i) \,| p \rangle  
=   \lim_{x \rightarrow 0}  \frac{1}{2} x G(x,Q^2) \;. 
\label{xi2}
\nonumber
\end{eqnarray}
The $x$- and $A$-dependence of  $F_2(A)/ F_2(D)$, 
calculated for  $\xi^2 = 0.09 - 0.12$~GeV$^2$,  is given in the 
left panel of Fig.~1. Comparison to a leading 
twist shadowing  parameterization~\cite{Eskola:1998df}     
is also shown. The right panel of Fig.~1 indicates
the power law nature of the nuclear 
modification to the structure functions.  
The physical gluon exchange leads to a high twist 
contribution to the longitudinal structure function $F_L$ and
enhances the ratio $R = \sigma_L/\sigma_T$.  
We emphasize that both leading twist~\cite{Brodsky:2002ue}  
and high twist shadowing~\cite{Qiu:2003vd} have their origin 
in the {\em final state} coherent scattering. This provides 
a natural  explanation of the apparent {\em lack} of gluon 
shadowing in the NLO global analysis~\cite{deFlorian:2003qf} 
which is the only one directly sensitive to gluons.

\subsection{Neutrino-nucleus scattering} 

Neutrino-nucleus scattering 
provides the unique opportunity to 
separately study the effect of coherent power corrections for 
sea and valance quarks~\cite{Qiu:2004qk} through the structure
functions:   
\begin{eqnarray} 
 F_{1(3)}^{\nu A} (x_B,Q^2) & \approx & A (2) 
    \left[  \sum_{D,U} |V_{DU}|^2 
\phi_D^A \left(x_B + x_B \frac{\xi^2 (A^{1/3}-1)}{Q^2} 
+ x_B \frac{M_U^2}{Q^2},Q^2\right)   \right.
 \nonumber   \\
 && \left.  +(-)  \sum_{\bar{U},\bar{D}} |V_{\bar{U} \bar{D}}|^2 
\phi_{\bar{U}}^A \left(x_B +   x_B \frac{\xi^2 (A^{1/3}-1)}{Q^2}  
 + x_B \frac{M_D^2}{Q^2} ,Q^2 \right)   \right] \, .  \;\;
\label{F13resWp}  
\end{eqnarray}
Here $V_{DU}$ are the CKM matrix elements.   
Eq.~(\ref{F13resWp}) identifies the nuclear enhanced high
twist corrections with dynamical mass 
$m_{dyn}^2 = \xi^2 (A^{1/3}-1)$ generated by the final state
parton scattering through direct comparison to $M^2_{U,D}$.

The modification to the structure functions $F_2(x,Q^2)$ and 
$xF_3(x,Q^2)$ for two select values of $x_B$ are shown in the left
panel of Fig.~2. These give a good description of the observed
power law deviation of the reduced cross sections measured by 
NuTeV~\cite{Radescu:2004zn,Tzanov:2003gq} from the leading twist pQCD 
at small values of $Q^2$. Note the difference in the ``shadowing''
of $F_2$ and $x F_3$ due to  the different steepness of sea 
and valence quark PDFs (in $x$). 
The right panel of Fig.~2 demonstrates
the improved agreement between data and theory for the 
Gross-Llewellyn-Smith sum rule~\cite{Qiu:2004qk}: 
 \begin{equation}
\Delta_{\rm GLS} 
\equiv  \frac{1}{3}\left(3 - S_{\rm GLS}\right)
= \frac{\alpha_s(Q^2)}{\pi} + \frac{{\cal G}}{Q^2} 
+{\cal O}(Q^{-4}) \; .
\end{equation} 

\begin{center}

\begin{figure}[!t]
  \includegraphics[height=2.5in]{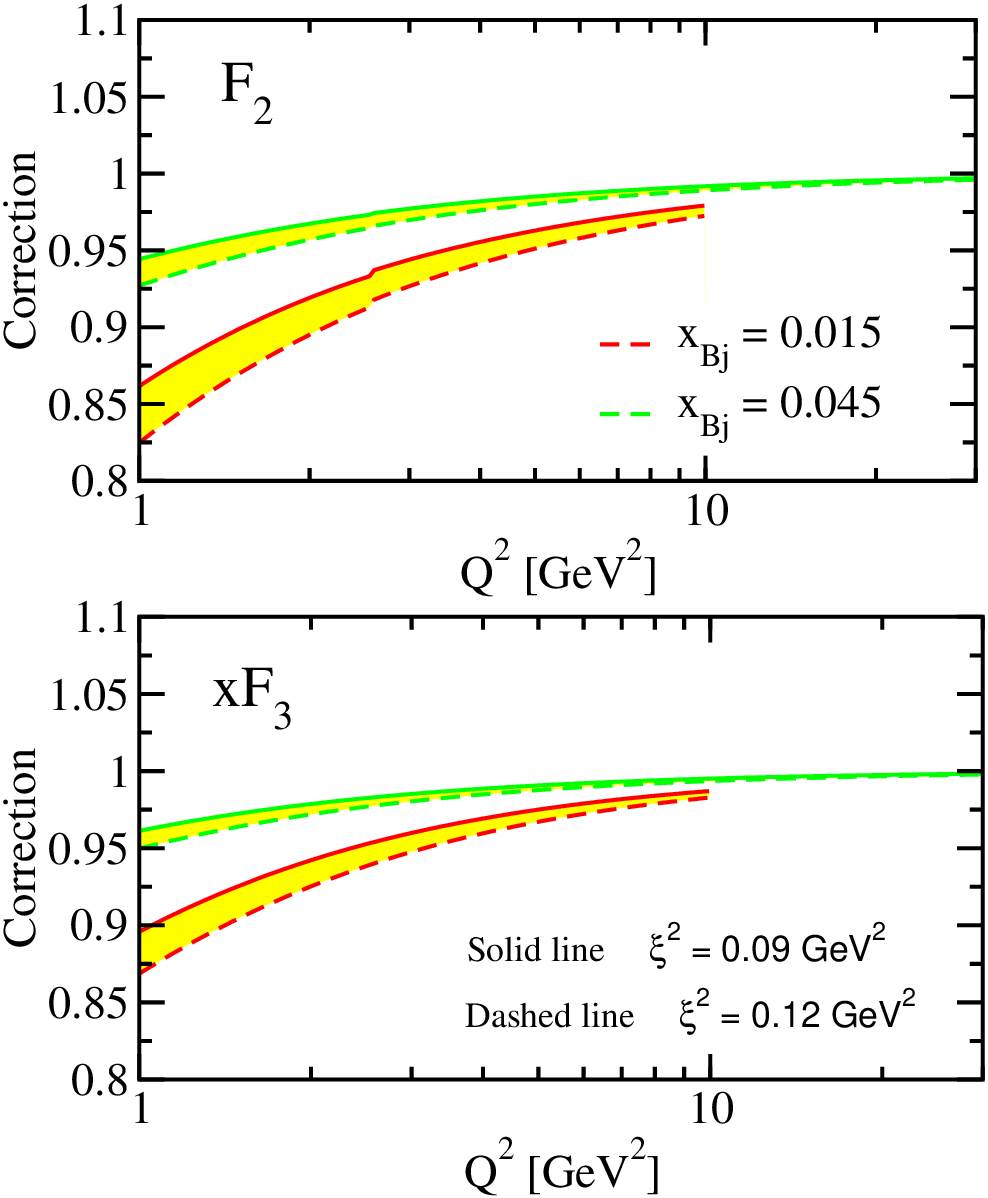} \hspace*{1cm}
  \includegraphics[height=1.8in]{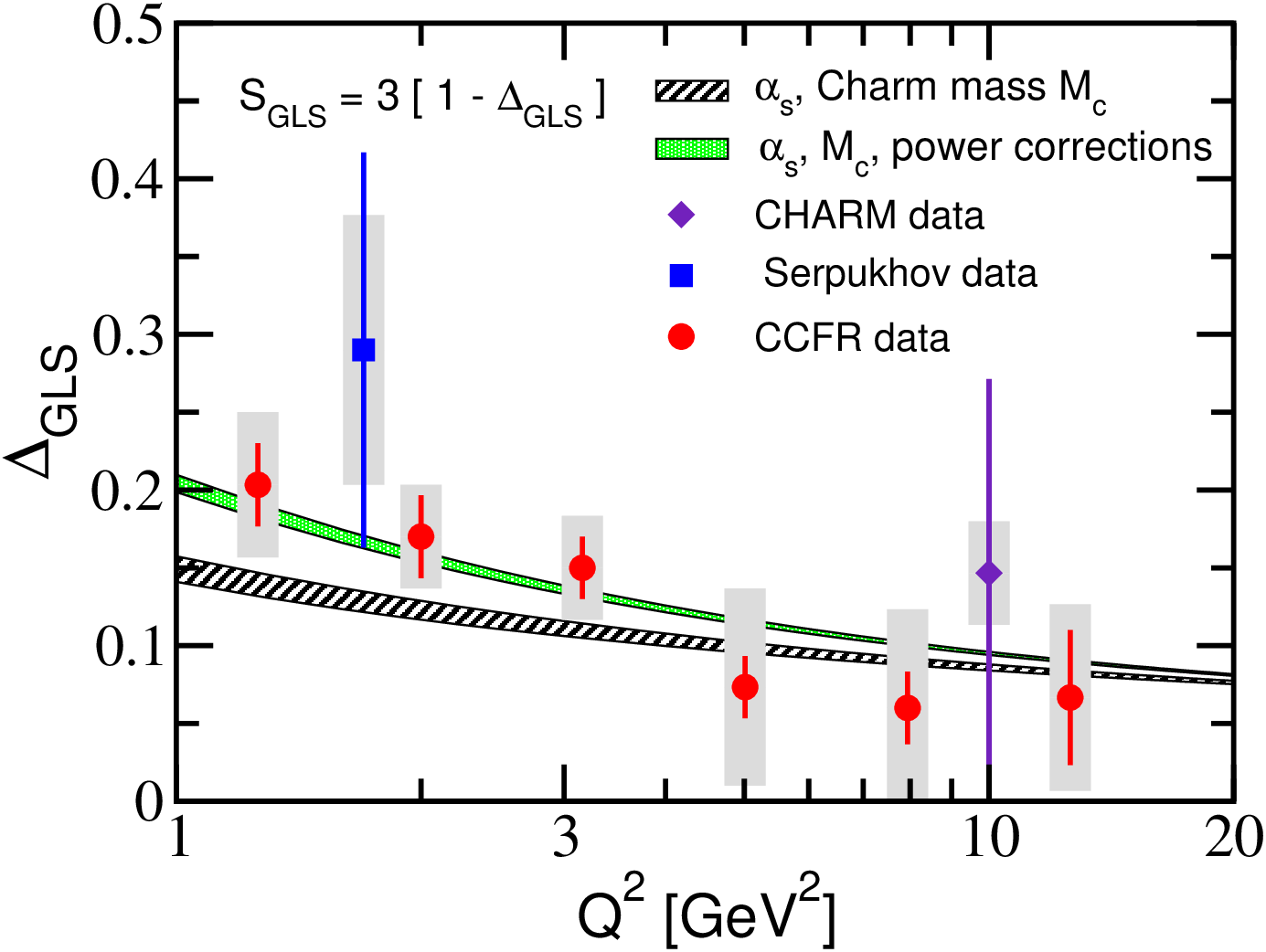}
  \caption{Left panel: power corrections to the 
structure functions $F_2(x,Q^2)$ and 
$xF_3(x,Q^2)$~\cite{Qiu:2004qk} for two values of $x_B$  
corresponding
to NuTeV measurements~\cite{Tzanov:2003gq}.
Right panel:  high twist 
modification to the Gross-Llewellyn-Smith 
sum rule $\Delta_{GLS}$~\cite{Qiu:2004qk}.  }
\end{figure}

\end{center}

\subsection{Proton-nucleus collisions}

The $p+A$ analogue of the DIS coherent power corrections
is the final state interactions of the small $x_b$ parton in 
the  $|\hat{t}| \ll |\hat{s}|, |\hat{u}|$ regime. Here 
$\hat{t} = q^2=(x_aP_a-P_c/z_1)^2$  and the $x_b$ rescaling 
in the  lowest order pQCD formalism 
reads~\cite{Qiu:2004da}:
\begin{eqnarray}
 F_{ab\rightarrow cd}(x_b) & \Rightarrow &
  F_{ab\rightarrow cd}\left(x_b \left[ 1+ C_d
\frac{\xi^2}{-t} (A^{1/3}-1) \right]\right)  \;.
\label{resum}
\end{eqnarray} 
In Eq.(\ref{resum}) $F_{ab\rightarrow cd}(x_b) = 
|M_{ab\rightarrow cd}|^2 \phi(x_b)/x_b$ and $C_d$ is a color factor, 
$C_{q(\bar{q})}=1$ and
$C_g=C_A/C_F=9/4$ for quark (antiquark) and gluon, respectively.

The left panel of Fig.~3 shows the {\em upper limit} on the centrality 
and rapidity dependent suppression $R^{(1)}_{pA}$ of single 
inclusive hadron production at RHIC. Data is from BRAHMS~\cite{Arsene:2004ux}. 
Additional nuclear suppression arises form the energy loss in 
cold nuclei~\cite{Kopeliovich:2005ym}. The right panel shows 
the suppression of away side dihadron correlations   
$R^{(2)}_{pA}$ versus transverse momentum, rapidity and centrality 
on the example of the area of the correlation function 
$C(\Delta \phi)= dN^{h_1,h_2}/d \Delta \phi$ 
The pronounced $p_{T_2}$ dependence is consistent with 
STAR preliminary data~\cite{Ogawa:2004sy}.

\begin{center}

\begin{figure}[!t]
  \includegraphics[height=3in]{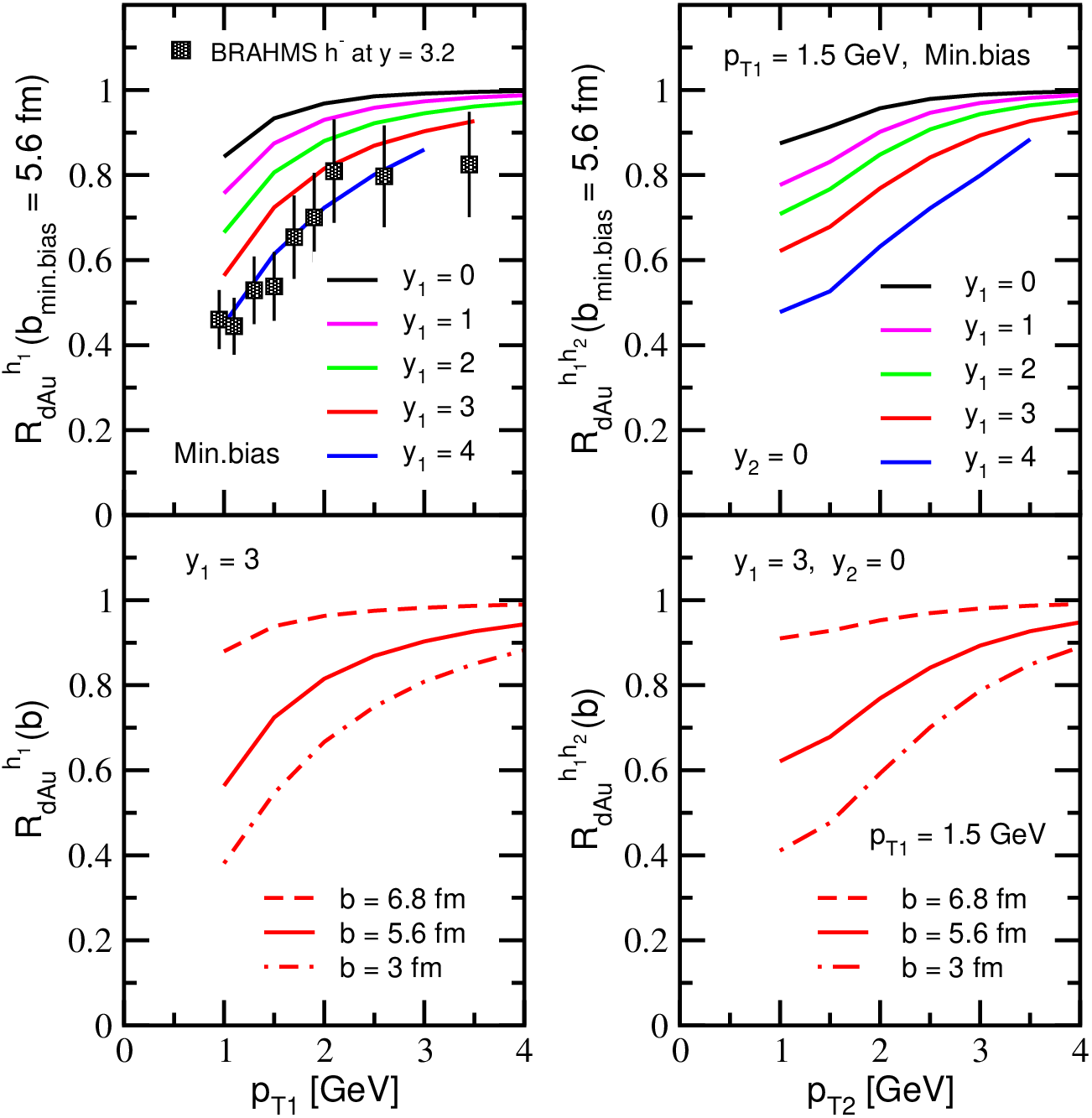} \hspace*{1cm}
  \includegraphics[height=2.5in]{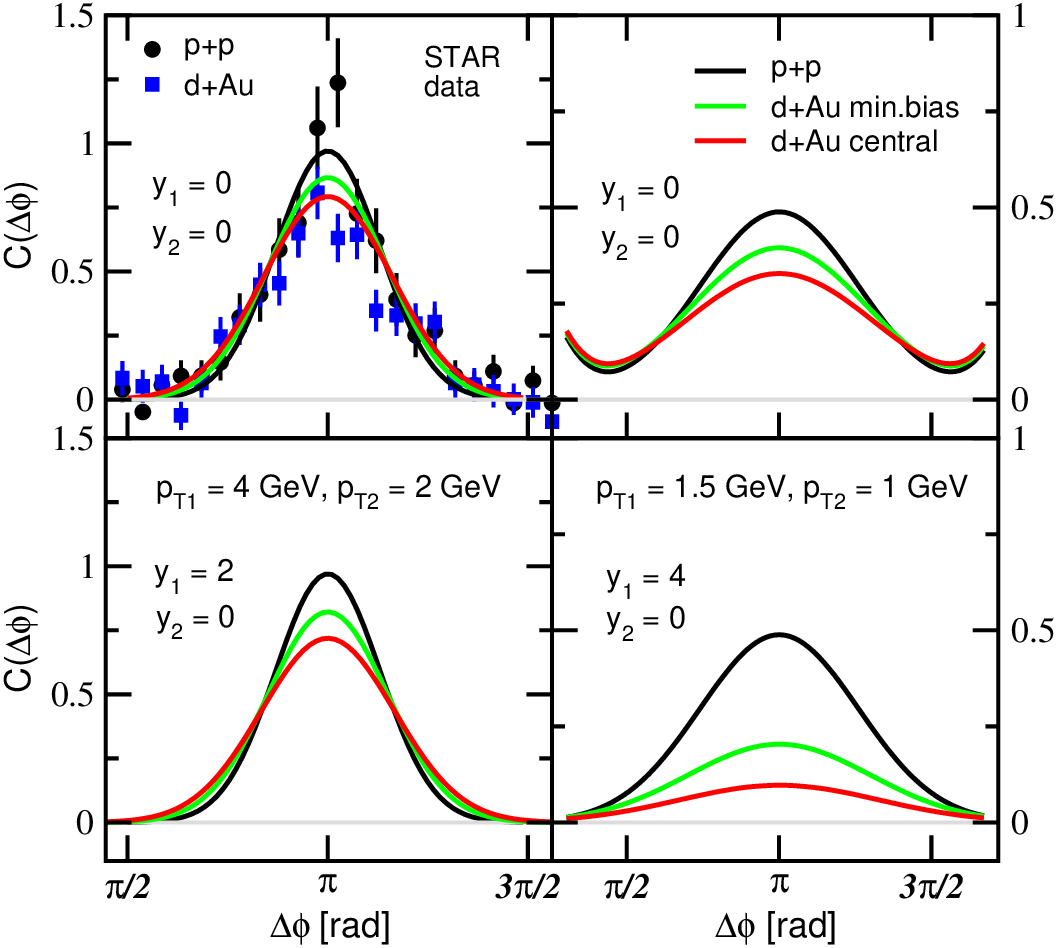}
  \caption{Left panel: upper limit on the suppression of the
single inclusive particle production $R^{(1)}_{pA}(p_{T_1})$ 
from  coherent power corrections versus rapidity 
and centrality~\cite{Qiu:2004da}. 
Data is from  BRAHMS~\cite{Arsene:2004ux}. 
Right panel: suppression of the 
double inclusive cross section 
$R^{(2)}_{pA}(p_{T_1},p_{T_2})$ for different rapidity gaps,
$p_{T_1},p_{T_2}$ ranges and centrality. }

\end{figure}

\end{center}


{\bf Acknowledgments:} 
Useful discussion with S.~J.~Brodsky, R.~Jaffe, J. W.~Qiu 
and M.~Tzanov is acknowledged. This work is supported
by the J.~Robert Oppenheimer Fellowship of the Los Alamos 
National Laboratory and by the US
Department of Energy.

\bibliographystyle{aipproc}   

\end{document}